\documentclass{PoS}
\let\ifpdf\relax
\usepackage{url}
\usepackage{cite}
\usepackage{lineno}
\newcommand{\Dy}{\relax}

\title{Measurements of the top quark mass using the ATLAS and CMS
  detectors at the LHC}

\ShortTitle{Measurements of the top quark mass using the ATLAS and CMS
  detectors at the LHC}

\author{\speaker{Sven Menke}%
  \thanks{On behalf of the ATLAS and CMS collaborations.}\\%
  Max-Planck-Institut f{\"u}r Physik\\%
  F{\"o}hringer Ring 6\\%
  80805 M{\"unchen}, Germany\\%
  E-mail: \email{menke@mppmu.mpg.de}}

\abstract{The latest measurements of the top quark mass obtained by
  the ATLAS and CMS experiments at the LHC for centre-of-mass energies
  of $7$ and $8\,$TeV are presented. The mass of the top quark is
  measured using several methods and channels, including the
  reconstructed invariant mass distribution of the top quark and
  shapes of top quark decay distributions. Measurements based on the
  inclusive ${\rm t}\bar{\rm t}$ cross section and novel observables
  based on the differential cross section in the ${\rm t}\bar{\rm
    t}+1\textrm{jet}$ channel are also discussed. The results of the
  various channels are combined and compared to the world average.}

\FullConference{XXIII International Workshop on Deep-Inelastic Scattering\\%
		 27 April - May 1 2015\\%
		 Dallas, Texas}

\begin{document}
%
\section{Introduction}
\label{sec:Intro}
The top quark is by far the heaviest known fermion and the heaviest
known fundamental particle. This gives the top-quark mass a unique
role in over-constraining Standard Model (SM) fits~\cite{art:GFitter}
and testing their validity in comparisons to direct mass
measurements. Together with the mass of the Higgs boson the top-quark
mass has consequences on the SM vacuum
stability~\cite{art:Vacuum_stability}.
  
With integrated luminosities of about $5\,{\rm fb}^{-1}$ and $20\,{\rm
  fb}^{-1}$ for both LHC experiments (ATLAS~\cite{art:ATLAS} and
CMS~\cite{art:CMS}) at $7$ and $8\,{\rm TeV}$, respectively, the
statistical and systematic uncertainties on the top-quark mass reach
levels well below $1\,{\rm GeV}$ -- with smaller uncertainties reached
at $8\,{\rm TeV}$ due to the increased statistics. The distinction of
the theoretical description of the measured parameter -- either the
parameter in the underlying Monte Carlo generator, the mass term in
the top-quark propagator (the pole mass) or the mass in a well defined
low-scale short distance
scheme~\cite{art:Top_Mass_Hoang_2008,art:Top_Mass_Moch_2014} -- is
gaining in importance.
\section{Template and Ideogram Based Measurements}
The typical analysis path for measurements of the top-quark mass
relying on the mass parameter of the underlying Monte Carlo generator
is to reconstruct and select ${\rm t}\bar{\rm t}$ candidate events in
data and simulations -- often refined by a kinematic fit that
constrains the four-vectors of the decay products of the top-quark
candidates, within assigned uncertainties, to stem from a heavy quark
decay of the same mass for both candidates. The decay is assumed to
proceed via ${\rm t}\to{\rm W}{\rm b}$. The known ${\rm W}$ mass is
typically used in the two possible decay channels of the ${\rm W}$ to
two quark-jets or a lepton and neutrino to further constrain the
kinematic fit. In the end a likelihood fit of the reconstructed
top-quark mass for the hadronic decay of the ${\rm W}$ or estimators
sensitive to it like the invariant mass of the lepton and ${\rm b}$,
$m_{\ell{\rm b}}$, in the leptonic ${\rm W}$ decay channel with
$m_\textrm{\footnotesize top}$ as free parameter is used to measure
the top-quark mass. In ATLAS the likelihood fits are often based on
templates~\cite{art:Templates}, while CMS uses both templates and
ideograms~\cite{art:Ideograms}.  Templates are
probability-density-functions constructed from full Monte Carlo
simulations in the final observables (for example the reconstructed
top-quark mass). For a variety of different top-quark-mass settings in
the Monte Carlo and, optionally, variations in other quantities, like
a systematic shift in the jet-energy scale (JES or JSF), templates are
obtained for signal and background samples. Their shapes are
parameterised and for the signal samples the shape parameters are
expressed as polynomial in $m_\textrm{\footnotesize top}$ and the
other varied quantities. Ideograms extend the idea of templates by
allowing multiple permutations per event -- in the signal templates
and in the final observable. The signal templates for the
reconstructed top-quark mass can differ for example by the number of
correct matches of the reconstructed decay products with
generator-level partons and in each event all possible assignments of
reconstructed objects to partons are used with a weight proportional
to the goodness-of-fit probability ($p_\textrm{\footnotesize g.o.f}$).

\subsection{Lepton plus Jets Channel}
\begin{figure}[htb]
  \begin{center}
    \parbox[c]{0.54\textwidth}{%
      \resizebox{0.54\textwidth}{!}{\rotatebox{270}{%
          \includegraphics{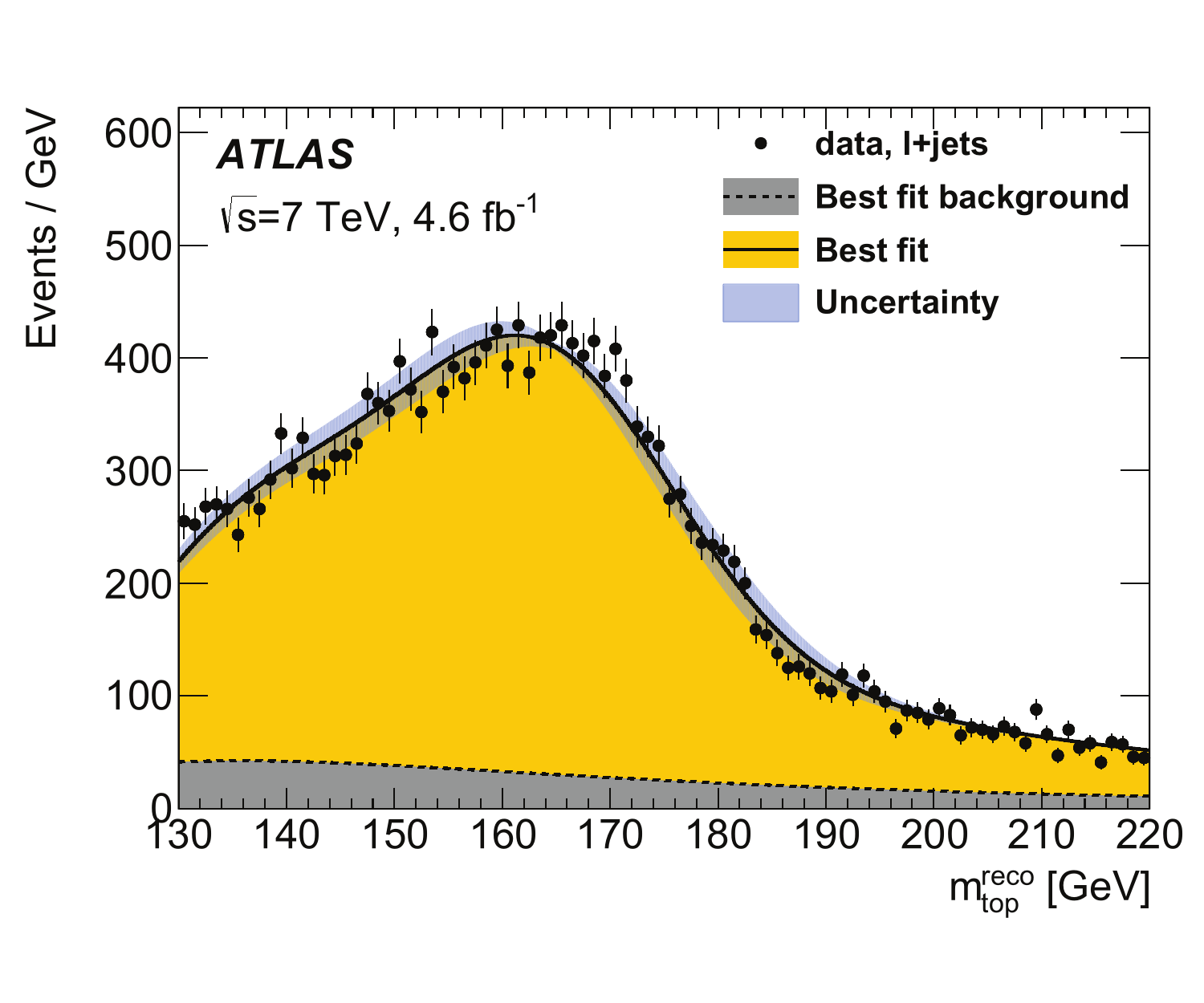}}}}\hfill%
    \parbox[c]{0.45\textwidth}{%
      \resizebox{0.45\textwidth}{!}{%
        \includegraphics{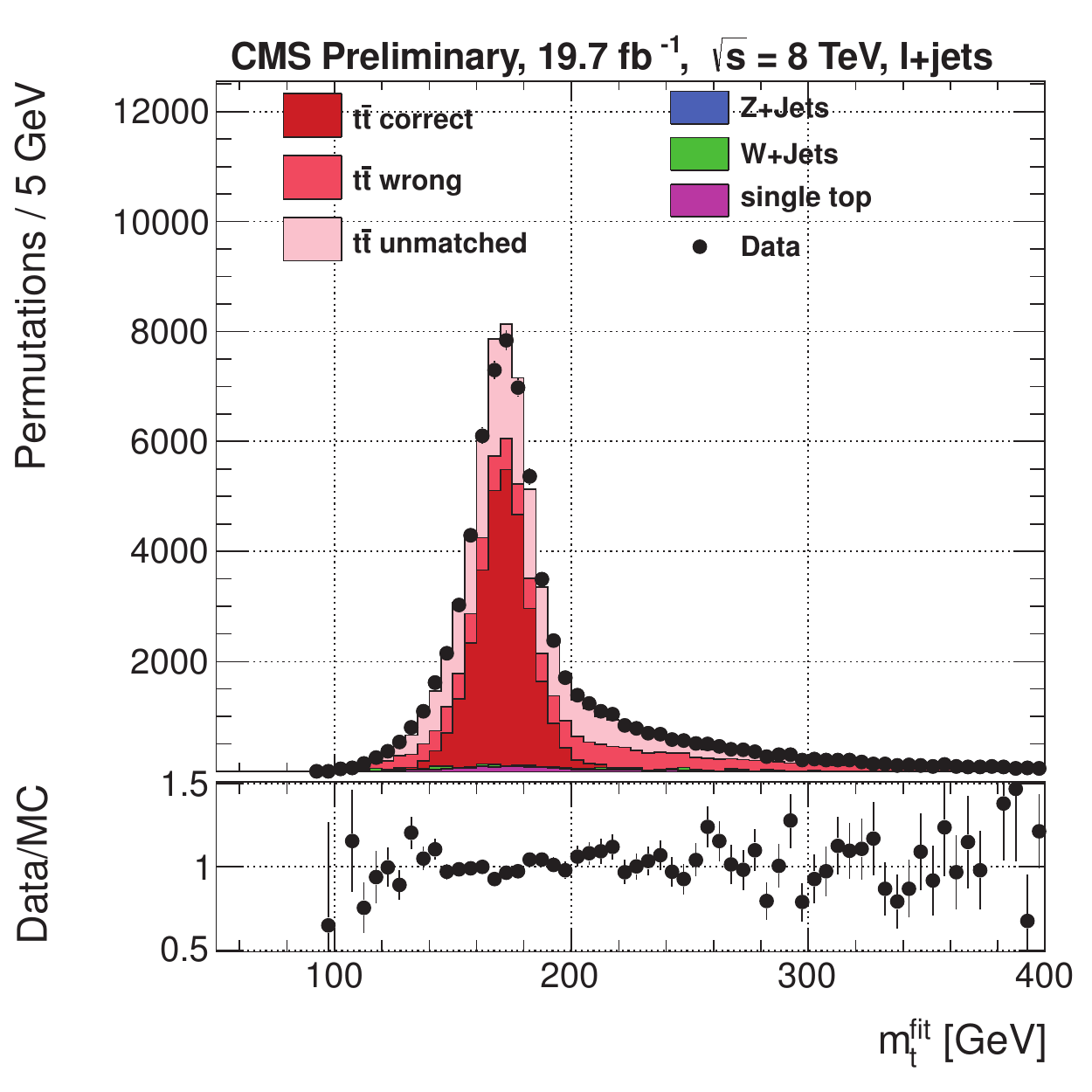}}}
    \caption{The reconstructed top-quark mass. Overlaid is the
      template fit for ATLAS~\cite{art:ATLAS_lj_ll_7TeV} on the left
      and the ideogram fit for CMS~\cite{art:CMS_lj_8TeV} on the
      right.}\label{fig:ATLAS_lj_mtop_CMS_lj_mtop}
  \end{center}
\end{figure}
Experimentally, the most precise measurements are achieved in the
lepton+jets channel, benefitting from moderate backgrounds, due to the
lepton requirements, and one fully reconstructible top-quark
candidate. The most recent analyses by
ATLAS~\cite{art:ATLAS_lj_ll_7TeV} (at $\sqrt{s}=7\,{\rm TeV}$) and
CMS~\cite{art:CMS_lj_8TeV} (at $\sqrt{s}=8\,{\rm TeV}$) use events
with isolated single electrons or muons with large transverse momentum
($p_\perp$) in the central detector and at least $4$ central jets with
large $p_\perp$. At least one (ATLAS) or exactly two (CMS) of the jets
need to be ${\rm b}$-tagged. Both experiments employ kinematic fits as
described above and the best permutation is retained in the ATLAS
analysis only, while all are kept in the CMS analysis with the
appropriate $p_\textrm{\footnotesize g.o.f}$. In ATLAS a template fit
is performed in three uncorrelated observables, the reconstructed
top-quark mass, $m_\textrm{\footnotesize top}^\textrm{\footnotesize
  reco}$, shown in Figure~\ref{fig:ATLAS_lj_mtop_CMS_lj_mtop} (left),
the ratio of the transverse momenta of ${\rm b}$-tagged over light
jets, $R_{\rm bq}^\textrm{\footnotesize reco}$, and the mass of the
$\rm W$, $m_{\rm W}^\textrm{\footnotesize reco}$, without the
kinematic constraint to its known value. The two additional
observables are sensitive to the JES of ${\rm b}$-tagged and light
jets, respectively, which reduces the systematic error on the obtained
top-quark mass substantially: $m_\textrm{\footnotesize top} = 172.33
\pm 0.75_\textrm{\footnotesize stat} \pm 1.02_\textrm{\footnotesize
  sys}(0.58_\textrm{\footnotesize JES} \oplus
0.50_\textrm{\footnotesize bTag} \oplus 0.32_\textrm{\footnotesize
  ISR/FSR} \oplus ...)\,{\rm GeV}$, with the dominant systematic
uncertainties stemming from JES, ${\rm b}$-tagging (bTAG) and the
modelling of initial- and final-state radiation (ISR/FSR).  In CMS the
ideograms are obtained in two uncorrelated observables, the top-quark
mass after kinematic fit for all $p_\textrm{\footnotesize
  g.o.f}$-weighted permutations, $m_{\rm t}^\textrm{\footnotesize
  fit}$, shown in Figure~\ref{fig:ATLAS_lj_mtop_CMS_lj_mtop} (right),
and the mass of the $\rm W$, $m_{\rm W}^\textrm{\footnotesize reco}$,
without the mass constraint but also $p_\textrm{\footnotesize
  g.o.f}$-weighted. The latter reduces the JSF systematic uncertainty
on the top-quark mass: $m_\textrm{\footnotesize top} = 172.04 \pm
0.19_\textrm{\footnotesize stat} \pm 0.75_\textrm{\footnotesize sys}
(0.41_\textrm{\footnotesize Flavour JSF} \oplus
0.27_\textrm{\footnotesize PileUp} \oplus 0.26_\textrm{\footnotesize
  JER} \oplus ...)\,{\rm GeV}$, with the dominant systematic
uncertainties stemming from JSF of $\rm b $-tagged jets (Flavour JSF),
the modelling of multiple soft pp interactions (PileUp) and the
modelling of the jet energy resolution (JER).
\subsection{Di-Lepton Channel}
The cleanest samples of ${\rm t}\bar{\rm t}$ events are obtained in
the di-lepton channel, by requiring exactly two oppositely charged
leptons (${\rm ee}$, ${\mu\mu}$, or ${\rm e}\mu$), with mass-vetoes
against $\rm Z$ and lower mass neutral states in the same flavour
channels, large missing transverse momentum, and at least two
jets. CMS~\cite{art:CMS_ll_8TeV} (at $\sqrt{s}=8\,{\rm TeV}$) keeps
the two $\rm b$-tagged jets leading in $p_\perp$ or supplements with
the leading un-tagged jet.  ATLAS~\cite{art:ATLAS_lj_ll_7TeV} (at
$\sqrt{s}=7\,{\rm TeV}$) requires exactly one or two $\rm b$-tags and
assigns the role of the second ${\rm b}$ to the one with the largest
${\rm b}$-tag weight in the one $\rm b$-tag case.  CMS constructs a
top-quark-mass estimator $m_\textrm{\footnotesize peak}$ from $500$
randomised re-reconstructions within the assigned two- and four-vector
uncertainties per event as the one with the highest leading-order
matrix-element weight. A quadratic fit to the log likelihood values
obtained from signal and background template fits at fixed generator
mass points leads after un-blinding to: $m_\textrm{\footnotesize top}
= 172.47 \pm 0.17_\textrm{\footnotesize stat} \pm
1.40_\textrm{\footnotesize sys} (0.87_{\mu_{R,F}} \oplus
0.67_\textrm{\footnotesize b-frag} \oplus 0.61_\textrm{\footnotesize
  JES} \oplus ...)\,{\rm GeV}$. Renormalisation and factorisation
scale variations ($\mu_{R,F}$), the modelling of $\rm b$ fragmentation
(b-frag), and the JES dominate the systematic uncertainty.  An
alternative blinded template fit to $m_{\ell{\rm b}}$ in the ${\rm
  e}\mu$ channel leads to a comparable result~\cite{art:CMS_mlb_8TeV}:
$m_\textrm{\footnotesize top} = 172.2 \pm 1.3\,{\rm GeV}$.  The ATLAS
fit to $m_{\ell{\rm b}}$ signal and background templates, where the
permutation with the lowest average $m_{\ell{\rm b}}$ is retained,
yields: $m_\textrm{\footnotesize top} = 173.79 \pm
0.54_\textrm{\footnotesize stat} \pm 1.30_\textrm{\footnotesize sys}
(0.75_\textrm{\footnotesize JES} \oplus 0.68_\textrm{\footnotesize
  bJES} \oplus 0.53_\textrm{\footnotesize hadro.} \oplus ...)\,{\rm
  GeV}$.  Without the additional constraints on JES and bJES these two
sources dominate the systematic uncertainty followed by the modelling
of hadronisation (hadro.). Since the correlation to the lepton+jets
result is $-7\%$ only, both are combined, yielding the preliminary
ATLAS Run-1 summary value~\cite{art:ATLAS_lj_ll_7TeV} given in
Section~\ref{sec:Conclusions}.
\subsection{All-Hadronic Channel}
The all-hadronic decay channel provides a fully reconstructed final
state for ${\rm t}\bar{\rm t}$ events but suffers from the large
multijet background from other QCD processes due to the absence of
isolated leptons. The only handle to suppress this background is ${\rm
  b}$-tagging and, in addition, data-driven methods are needed to
estimate it.  ATLAS~\cite{art:ATLAS_had_7TeV} (at $\sqrt{s}=7\,{\rm
  TeV}$) requires exactly two ${\rm b}$-tags among the leading $4$
jets and at least $6$ central jets, $5$ with $p_\perp > 55\,{\rm GeV}$
and $p_\perp > 30\,{\rm GeV}$ for the $6^\textrm{\footnotesize
  th}$. The background is estimated from control regions defined by
the number of ${\rm b}$-tags and $2$ bins in the $p_\perp$ of the
$6^\textrm{\footnotesize th}$ jet.  CMS~\cite{art:CMS_had_8TeV} (at
$\sqrt{s}=8\,{\rm TeV}$) requires at least $6$ central jets with
$p_\perp>60\,{\rm GeV}$ for the leading $4$ and $p_\perp>30\,{\rm
  GeV}$ for the other $2$ and exactly $2$ ${\rm b}$-tags among the
leading $6$.  The multijet background is derived with an event-mixing
technique.  Both experiments reject events not compatible with the
${\rm t}\bar{\rm t}$ hypothesis. ATLAS fits the ratios of $3$-jet mass
over $2$-jet mass, $R_{3/2}$, using the permutation of the six jets
that fits the assumed ${\rm t}\bar{\rm t}$ to $6$ partons hypothesis
best, to signal and background templates: $m_\textrm{\footnotesize
  top} = 175.1 \pm 1.4_\textrm{\footnotesize stat} \pm
1.2_\textrm{\footnotesize sys} (0.62_\textrm{\footnotesize bJES}
\oplus 0.51_\textrm{\footnotesize JES} \oplus
0.50_\textrm{\footnotesize hadro.} \oplus ...)\,{\rm GeV}$. While
still being statistically limited, the dominant systematic uncertainty
stems from bJES, followed by JES and the modelling of
hadronisation. The statistical error takes the $\sim60\%$ correlation
of the two measured $R_{3/2}$ in each event into account.  Like
in~\cite{art:CMS_lj_8TeV} the CMS analysis uses $m_{\rm
  t}^\textrm{\footnotesize fit}$ and $m_{\rm W}^\textrm{\footnotesize
  reco}$ in an ideogram based fit with $m_\textrm{\footnotesize top}$,
JSF, signal fraction and fraction of correctly assigned permutations
as free parameters: $m_\textrm{\footnotesize top} = 172.08 \pm
0.36_\textrm{\footnotesize stat} \pm 0.83_\textrm{\footnotesize sys}
(0.36_\textrm{\footnotesize Flavour JSF} \oplus
0.31_\textrm{\footnotesize PileUp} \oplus 0.28_\textrm{\footnotesize
  JES} \oplus ...)\,{\rm GeV}$. Flavour JSF and PileUp dominate the
systematics, followed by JES.  In both experiments the results are
cross-checked with several thousand pseudo-experiments and corrected
for small observed biases.
\section{Measurements of the pole mass}
In contrast to the template/ideogram methods discussed above,
cross-section-like observables can be used to compare QCD predictions
depending on the pole mass, $m_\textrm{\footnotesize
  top}^\textrm{\footnotesize pole}$, with unfolded data. The unfolding
removes detector effects. The advantage lies in the larger
independence from the mass definition in Monte Carlo generators. For
the total cross-section, however, a $5\%$ uncertainty translates into
a $1\%$ uncertainty in the top-quark mass~\cite{art:xsec_shape} and
the difference from NLO to NNLO is large ($\sim10\%$).  Experimentally
the challenges lie in the unfolding of data and the absolute
normalization. New shape-like observables as proposed
in~\cite{art:new_obs} and discussed in Section~\ref{sec:ttj} reduce
both theoretical and experimental uncertainties.
\begin{figure}[htb]
  \begin{center}
    \parbox[c]{0.49\textwidth}{%
      \resizebox{0.49\textwidth}{!}{%
          \includegraphics{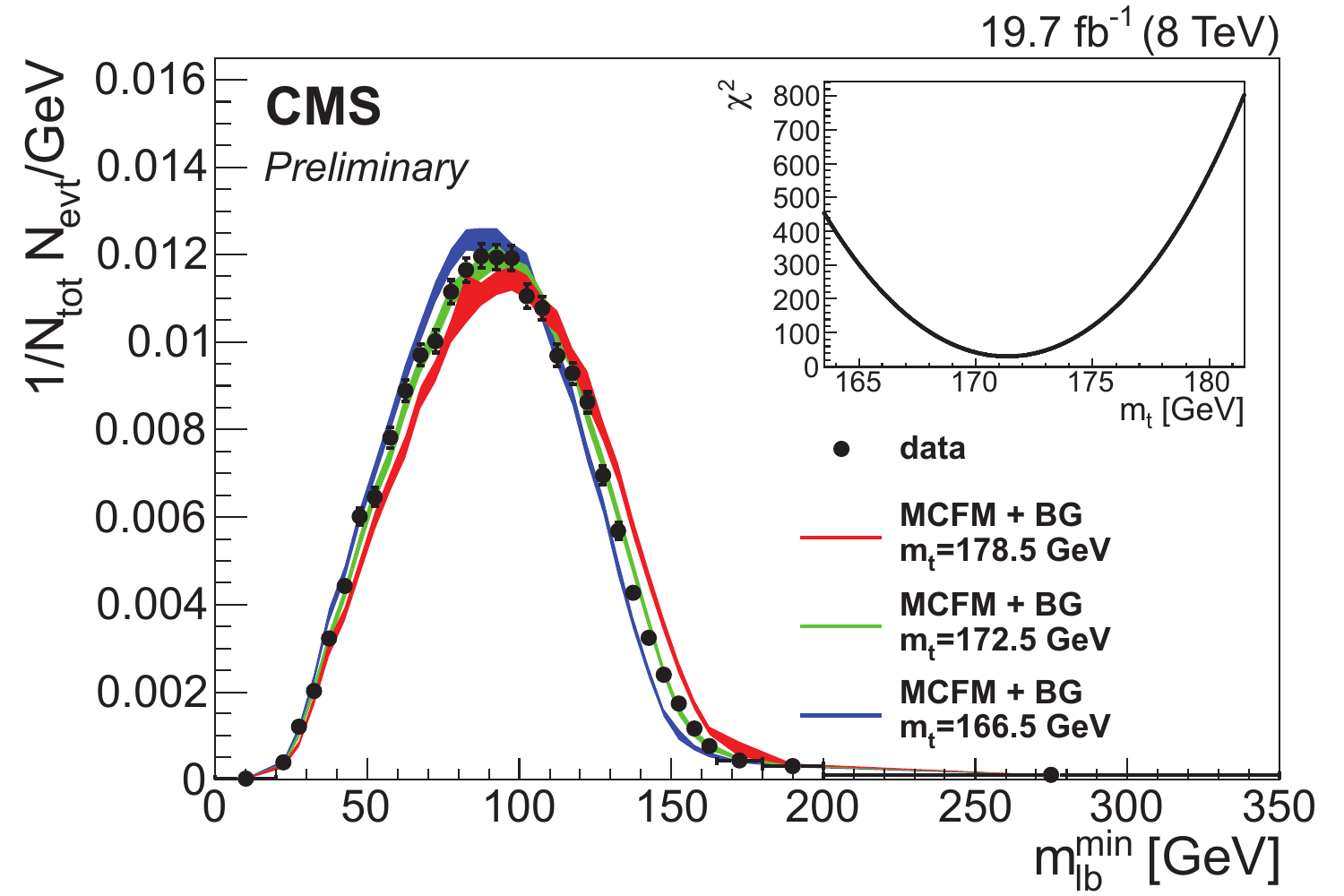}}}\hfill%
    \parbox[c]{0.49\textwidth}{%
      \resizebox{0.49\textwidth}{!}{%
        \includegraphics{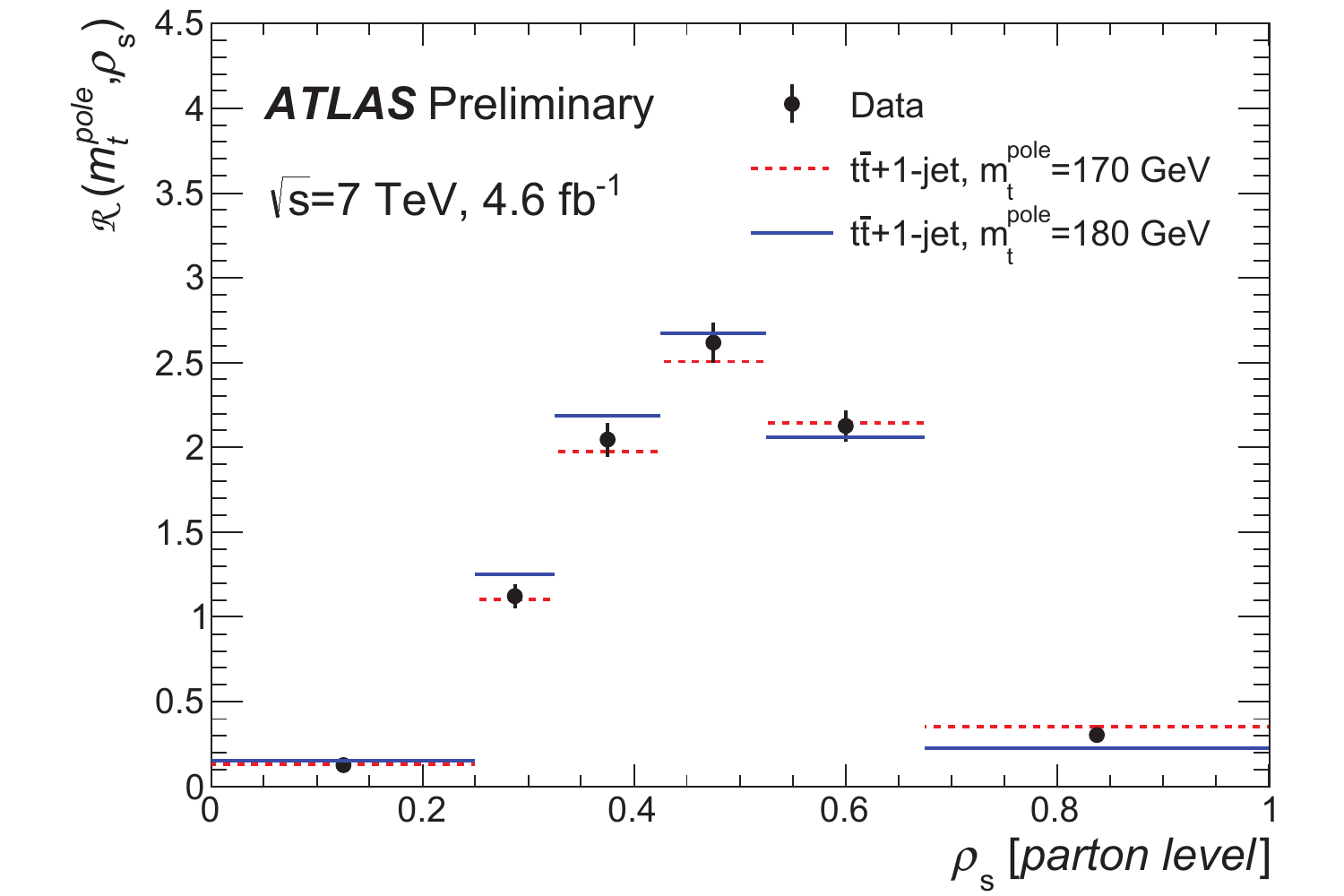}}}
    \caption{The shape of $m_{\ell{\rm b}}^\textrm{\footnotesize min}$
      from CMS~\cite{art:CMS_mlb_8TeV} on the left and the unfolded
      ${\cal R}(m_\textrm{\footnotesize top}^\textrm{\footnotesize
        pole},\rho_s)$ distribution as measured by
      ATLAS~\cite{art:ATLAS_ttj_7TeV} on the right. Theoretical,
      pole-mass driven predictions are overlaid for
      both.}\label{fig:ATLAS_ttj_CMS_mlb}
  \end{center}
\end{figure}
\subsection{Di-Lepton Channel}
The ${\rm t}\bar{\rm t}$ production cross-section in the di-lepton
channel is measured by ATLAS~\cite{art:ATLAS_xsec_7_8TeV} at
$\sqrt{s}=7$ and $8\,{\rm TeV}$ and by CMS~\cite{art:CMS_xsec_7TeV} at
$\sqrt{s}=7\,{\rm TeV}$. Both experiments assume a top-quark mass of
$m_\textrm{\footnotesize top}^\textrm{\footnotesize MC} = 172.5\,{\rm
  GeV}$ in simulations to extract the reconstruction efficiency.
Comparing the measured cross-section with theory predictions can be
used to extract the top-quark pole mass.  A small residual bias from
the assumed top-quark mass in the Monte Carlo (through single-top
background and reconstruction efficiencies) is corrected for.  The CMS
analysis~\cite{art:CMS_pole_7TeV} from the cross-section (fixing
$\alpha_{\rm s}(m_{\rm Z})$ to $0.118$) gives:
$m_\textrm{\footnotesize top}^\textrm{\footnotesize pole} =
176.7{{+3.0}\atop{-2.8}}({{+2.1}\atop{-2.0}}_\textrm{\footnotesize
  meas.~xsec} \oplus {{+1.5}\atop{-1.3}}_\textrm{\footnotesize PDF}
\oplus 0.9_{\mu_{R,F}} \oplus 0.9_{E_\textrm{\footnotesize LHC}}
\oplus ...)\,{\rm GeV}$, with larger experimental errors (meas.~xsec)
compared to the ideogram results and sizeable theoretical and
additional systematic uncertainties (from parton density functions
(PDF), $\mu_{R,F}$, and the energy uncertainty of the LHC machine
($E_\textrm{\footnotesize LHC}$)).  In the ATLAS
analysis~\cite{art:ATLAS_xsec_7_8TeV} theoretical uncertainties
dominate leading to a large correlation of the results for $\sqrt{s} =
7$ and $8\,{\rm TeV}$. For $8\,{\rm TeV}$ the result is:
$m_\textrm{\footnotesize top}^\textrm{\footnotesize pole} =
174.1\pm0.3_\textrm{\footnotesize stat}\pm2.6_\textrm{\footnotesize
  sys+theo}(1.7_{\textrm{\footnotesize PDF}+\alpha_{\rm
    s}}\oplus{{+0.9}\atop{-1.3}}_{\mu_{R,F}} \oplus
1.2_\textrm{\footnotesize lumi} \oplus ...)\,{\rm GeV}$. Like for CMS
the PDF and $\mu_{R,F}$ uncertainties dominate, followed by the
uncertainty in the integrated luminosity (lumi).

Also at $\sqrt{s}=8\,{\rm TeV}$ CMS~\cite{art:CMS_mlb_8TeV} uses a
folding technique to map fixed order QCD calculations depending on the
top-quark pole mass as implemented in MCFM~\cite{art:MCFM} to predict
the shape in $m_{\ell{\rm b}}^\textrm{\footnotesize min}$. Here, the
combination yielding the smallest $m_{\ell{\rm b}}$ in the event is
kept, and referred to as $m_{\ell{\rm b}}^\textrm{\footnotesize min}$,
which is shown in Figure~\ref{fig:ATLAS_ttj_CMS_mlb} (left). The
response matrices in $m_{\ell{\rm b}}^\textrm{\footnotesize min}$ are
obtained from fully simulated {\sc
  MadGraph+Pythia+Geant4}~\cite{art:MadGraph,art:Pythia6,art:Geant4}
events. This approach leads to: $m_\textrm{\footnotesize
  top}^\textrm{\footnotesize pole} = 171.4\pm0.4_\textrm{\footnotesize
  stat}\pm1.0_\textrm{\footnotesize sys} (0.5_{\mu_{R,F}} \oplus
0.43_\textrm{\footnotesize JES} \oplus 0.43_\textrm{\footnotesize b
  frag} \oplus ...)\,{\rm GeV}$.  This result can be compared to the
mass extraction from the same data set via the total cross-section
calculated at NNLO: $m_\textrm{\footnotesize
  top}^\textrm{\footnotesize pole} = 173.7\pm0.3_\textrm{\footnotesize
  stat}\pm3.4_\textrm{\footnotesize sys} (1.3_\textrm{\footnotesize
  lumi}\oplus1.2_\textrm{\footnotesize
  bkgd}\oplus1.1_\textrm{\footnotesize ME} \oplus ...)\,{\rm GeV}$,
with large uncertainties stemming from the luminosity measurement,
background modeling (bkgd) and the assumed matrix element (ME). The
comparison demonstrates the advantage of shape-based over total
cross-section based methods.
\subsection{Top-Quark Pair plus Jet Channel}\label{sec:ttj}
A new type of differential cross section observable is suggested
in~\cite{art:new_obs} to measure the top-quark pole mass in the ${\rm
  t}\bar{\rm t}+1{\rm jet}$ channel: ${\cal R}(m_\textrm{\footnotesize
  top}^\textrm{\footnotesize pole},\rho_s) = \frac{\Dy 1}{\Dy
  \sigma_{{\rm t}\bar{\rm t}+1\textrm{\footnotesize jet}}}\frac{\Dy
  {\rm d}\sigma_{{\rm t}\bar{\rm t}+1\textrm{\footnotesize
      jet}}}{\Dy{\rm d}\rho_s}(m_\textrm{\footnotesize
  top}^\textrm{\footnotesize pole},\rho_s),$ where the differential is
taken in $\rho_s = 2m_0/\sqrt{s_{{\rm t}\bar{\rm t}j}}$, the ratio of
an arbitrary mass scale in the vicinity of $m_\textrm{\footnotesize
  top}$, here set to $m_0 = 170\,{\rm GeV}$ over the invariant ${\rm
  t}\bar{\rm t}+1{\rm jet}$ mass.  ATLAS~\cite{art:ATLAS_ttj_7TeV}
first selects ${\rm t}\bar{\rm t}$ candidate events at
$\sqrt{s}=7\,{\rm TeV}$ similar to the standard analysis in the
lepton+jets channel~\cite{art:ATLAS_lj_ll_7TeV} and an additional
central jet with $p_\perp > 50\,{\rm GeV}$. An SVD
unfolding~\cite{art:SVD} with a response matrix from {\sc
  Powheg+Pythia+Geant4}~\cite{art:Powheg,art:Pythia6,art:Geant4} maps
the measured $\rho_s$ to parton level. The unfolded distribution in
$\rho_s$ is shown in Figure~\ref{fig:ATLAS_ttj_CMS_mlb} (right).  The
pole mass is then obtained in a $\chi^2$-fit to $0.25 < \rho_s < 1$
with the last bin $\rho_s>0.675$ being the most sensitive one:
$m_\textrm{\footnotesize top}^\textrm{\footnotesize pole} = 173.7 \pm
1.5_\textrm{\footnotesize
  stat}{{+1.0}\atop{-0.5}}_\textrm{\footnotesize theo} \pm
1.4_\textrm{\footnotesize sys} (0.9_\textrm{\footnotesize JES+bJES}
\oplus 0.7_\textrm{\footnotesize ISR/FSR} \oplus
0.5_\textrm{\footnotesize PDF} \oplus ...)\,{\rm GeV}$.
\section{Conclusions}%
\label{sec:Conclusions}
The ATLAS and CMS experiments both measured the top-quark mass in a
variety of channels and with different methods in ${\rm p}{\rm p}$
collisions at centre-of-mass energies of $\sqrt{s}=7$ and $8\,{\rm
  TeV}$ (Run 1). Together with the Tevatron experiments D0 and CDF the
2014 world average of $m_\textrm{\footnotesize top} = 173.34 \pm
0.27_\textrm{\footnotesize stat} \pm 0.71_\textrm{\footnotesize
  sys}\,{\rm GeV}$~\cite{art:TeV_LHC_Comb_2014} was published. Since
then more results became public leading to a preliminary Run 1 average
by ATLAS of $m_\textrm{\footnotesize top} = 172.99 \pm
0.48_\textrm{\footnotesize stat} \pm 0.78_\textrm{\footnotesize
  sys}\,{\rm GeV}$~\cite{art:ATLAS_lj_ll_7TeV} and a Run 1 average by
CMS of $m_\textrm{\footnotesize top} = 172.38 \pm
0.10_\textrm{\footnotesize stat} \pm 0.65_\textrm{\footnotesize
  sys}\,{\rm GeV}$~\cite{art:CMS_Run1_comb}. Within uncertainties the
results obtained from the different theoretical approaches (Monte
Carlo mass vs.~pole mass) agree well.
%
%
\section*{Acknowledgments}
 \label{Acknowledgments}
 I would like to thank the Top-Quark groups of ATLAS and CMS for providing
 me with the material presented here. 
\providecommand{\href}[2]{#2}\begingroup\raggedright\endgroup

\end{document}